\documentclass[twocolumn,preprintnumbers,amsmath,amssymb,jcp,nofootinbib]{revtex4}

\makeatletter
\def\@dotsep{4.5}
\makeatother

\usepackage{graphicx}% Include figre files
\usepackage{dcolumn}% Align table columns on decimal point
\usepackage{bm}% bold math
\usepackage{amsfonts}
\usepackage{amssymb}
\usepackage{amsmath}
\newcommand{\beq}{\begin{equation}}
\newcommand{\eeq}{\end{equation}}
\newcommand{\beqa}{\begin{eqnarray}}
\newcommand{\eeqa}{\end{eqnarray}}

\begin{document}

\title{Numerical determination of the exponents controlling the relationship between time, length and temperature in~glass-forming~liquids}

\author{Chiara Cammarota~$^{1,2}$}
\email{chiara.cammarota@roma1.infn.it}
%\affiliation{
%Dipartimento di Fisica, 'Sapienza' Universit\`a di Roma, 
%Piazzale Aldo Moro 5, 00185 Roma, Italy and 
%Center for Statistical Mechanics and Complexity (SMC), CNR - INFM}

%\affiliation{Center for Statistical Mechanics and Complexity (SMC), CNR - INFM}

\author{Andrea Cavagna~$^{2,3}$}
%\affiliation{
%Center for Statistical Mechanics and Complexity (SMC), CNR - INFM}
%\affiliation{Istituto Sistemi Complessi (ISC), CNR, via dei Taurini 19, 00185 Roma, Italy}

\author{Giacomo Gradenigo~$^4$}
\email{gradenigo@science.unitn.it}
%\affiliation{
%Dipartimento di Fisica and SOFT, Universit\`a di Trento, via Sommarive 14, 
%38050 Trento, Italy
%}

\author{Tomas S. Grigera~$^5$}
%\affiliation{Instituto de Investigaciones Fisicoqu{\'\i}micas
%  Te{\'o}ricas y Aplicadas (INIFTA -- CCT La Plata) and 
%  \\Departamento de F{\'\i}sica, Facultad de Ciencias Exactas,
%  Universidad Nacional de La Plata, c.c. 16, suc. 4, 1900 La Plata,
%  Argentina and \\Consejo Nacional de Investigaciones Cient{\'\i}ficas y
%  T{\'e}cnicas, Argentina.
%} 

\author{Paolo Verrocchio~$^4$}
%\affiliation{
%Dipartimento di Fisica and SOFT, University of Trento, via Sommarive 14, 
%38050 Trento, Italy
%}

\address{
$^1$~Dipartimento di Fisica, 'Sapienza' Universit\`a di Roma, 
Piazzale Aldo Moro 5, 00185 Roma, Italy\\ 
$^2$~Center for Statistical Mechanics and Complexity (SMC), CNR - INFM\\ 
$^3$~Istituto Sistemi Complessi (ISC), CNR, via dei Taurini 19, 00185 Roma, 
Italy\\ 
$^4$~Dipartimento di Fisica and SOFT, University of Trento, via Sommarive 14, 
38050 Trento, Italy\\ 
$^5$~Instituto de Investigaciones Fisicoqu{\'\i}micas
  Te{\'o}ricas y Aplicadas (INIFTA) and\\
  Departamento de F{\'\i}sica, Facultad de Ciencias Exactas,
  Universidad Nacional de La Plata, c.c. 16, suc. 4, 1900 La Plata,
  Argentina and\\CCT La Plata, Consejo Nacional de Investigaciones
  Cient{\'\i}ficas y T{\'e}cnicas, Argentina
}

\date{\today}% It is always \today, today,

%\preprint{preprint}

\begin{abstract}
There is a certain consensus that the very fast growth of the
relaxation time $\tau$ occurring in glass-forming liquids on lowering
the temperature must be due to the thermally activated rearrangement
of correlated regions of growing size. Even though measuring
the size of these regions has defied scientists for a while, there is
indeed recent evidence of a growing correlation length $\xi$ in
glass-formers. If we use Arrhenius law and make the mild assumption
that the free-energy barrier to rearrangement scales as some power
$\psi$ of the size of the correlated regions, we obtain a relationship
between time and length, $T \log \tau \sim \xi^\psi$. According to
both the Adam-Gibbs and the Random First Order theory the
correlation length grows as $\xi\sim (T-T_k)^{-1/(d-\theta)}$, even
though the two theories disagree on the value of $\theta$. Therefore,
the super-Arrhenius growth of the relaxation time with the temperature
is regulated by the two exponents $\psi$ and $\theta$ through the
relationship $T \log \tau \sim (T-T_k)^{-\psi/(d-\theta)}$. Despite a
few theoretical speculations, up to now there has been no experimental
determination of these two exponents. Here we measure them
numerically in a model glass-former, finding $\psi=1$ and $\theta=2$. 
Surprisingly, even
though the values we found disagree with most previous theoretical
suggestions, they give back the well-known VFT law for the relaxation
time, $T \log \tau \sim (T-T_k)^{-1}$.
\end{abstract}


\maketitle

\section{Introduction} \label{intro}

One of the most interesting open problems in the physics of
glass-forming liquids is how the slowing down of the
dynamics is related to the presence of a growing correlation
length. Even though the very steep growth of the relaxation time on
lowering the temperature is perhaps the experimentally most
conspicuous trait of glassy systems, the existence of an associated
growing length-scale has been a matter of pure speculation for quite a
long time. Yet, the qualitative idea that time grows because
relaxation must proceed through the rearrangement of larger and larger
correlated regions, is a sound one, so that many analytical, numerical
and experimental efforts have been devoted in the last fifteen years
to detect correlated regions of growing size in supercooled liquids.

The first breakthrough was to discover the existence of {\it dynamical} 
correlation length $\xi_d$. This is the spatial span of the
correlation of the mobility of particles and it is directly related to
the collective dynamical rearrangements of the system \cite{Ediger}. More
recently, a {\it static} correlation length $\xi$ has been measured,
by studying how deeply amorphous boundary conditions penetrate within
a system \cite{PRL,NP}. Both lengthscales 
grow when decreasing $T$,
and even though the relationship between dynamic and static
correlation length is still under investigation
\cite{FranzMontanari}, their existence strengthens the idea of a link
between time and length in glass-formers.

In principle, to uncover the formal nature of such link one simply
needs to compare the dependence of both relaxation time $\tau(T)$ and
correlation length $\xi(T)$ on the temperature, and infer the law
$\tau=\tau(\xi)$.  In practice, data on the correlation length are
still too scarce to pursue this indirect (parametric) road reliably.
Hence we must use a direct method to link length and time.

The super-Arrhenius growth of the relaxation time in fragile systems
suggests that the free-energy barrier $\Delta$ to relaxation must grow as well
on lowering $T$. It is therefore natural to assume that this 
barrier grows because the size of the regions to be rearranged, i.e.\ 
the correlation length $\xi$, does. A mild assumption is that
the following,
\begin{equation}
\Delta \sim \xi^{\psi},
\label{DvsRpsi}
\end{equation}
so that the Arrhenius law of activation gives
\begin{equation}
\tau \sim \exp \bigg( \frac{\xi^{\psi}}{T} \bigg)   .
\label{relaxation}
\end{equation}
Of course, this crucial equation is of little help as long the value
of the exponent $\psi$ remains unknown, but this is, unfortunately,
the current state of affairs.

To work out a more explicit relationship between time and
temperature, we need to know how precisely the correlation length
increases with decreasing $T$. This is, however, one of the most
disputed and open problems currently in the physics of glassy systems,
so that trying to follow a common path from now on is hopeless. We
choose to move in the framework of two theoretical schemes that,
despite many conceptual differences, share some common ground, namely
the Adam-Gibbs (AG) and the Random First Order (RFOT) theories. The two
common ideas of AG and RFOT are: 1) there exists a static correlation
length, whose growth is responsible for the growth of the relaxation
time, and 2) the static correlation length grows because the configurational 
entropy $S_c$ decreases. Although the two theories are really quite
different in explaining how and why the correlation length is connected 
to the configurational entropy, they both predict a power-law link,
\begin{equation}
\xi \sim \left( \frac{1}{S_c}\right)^\frac{1}{d-\theta},
\label{eq:1}
\end{equation}
where $d$ is the space dimension. The exponent $\theta$ is zero
according to AG (in fact, it is not even  
introduced in the theory),
\beq
\theta_\mathrm{AG}=0     ,
\eeq
whereas it has a crucial role within RFOT, where it regulates how
the interfacial energy of amorphous excitations grows with the size of the
excitations. The exponent $\theta$ is subject to the constraint
$\theta\leq d-1$. The value
\beq
\theta_\mathrm{RFOT} = d/2  
\eeq
for $\theta$ was proposed in \cite{KTW} from renormalization
group arguments.  Independently of the exact value of $\theta$, we see
that as long as $\theta >0$, the growth of $\xi$ with decreasing
configurational entropy predicted by RFOT is sharper than in AG.

From the behaviour of the (extrapolated) excess entropy of the
supercooled liquid with respect to the that of the crystal, it is
possible to see that the configurational 
entropy scales almost linearly with $T$,
\begin{equation}
S_c(T) \sim T-T_k     ,
\end{equation}
where $T_k$ is the so-called Kauzmann's temperature, or entropy crisis
point. In 
this way, the relationship between length and temperatures in the context of
the AG and RFOT schemes becomes
\beq
\xi \sim \left( \frac{1}{T-T_k}\right)^\frac{1}{d-\theta}   ,
\eeq
and that between time and temperature,
\beq
\tau \sim \exp \left[ \left(
\frac{A}{T-T_k}\right)^\frac{\psi}{d-\theta} \right]   ,
\label{girobatol}
\eeq 
where $A$ is a factor weakly dependent on temperature.

Hence, we see that the two exponents $\psi$ and $\theta$ contain all
the relevant information about the relationship between time, length
and temperature in supercooled liquids. However, little is known about
them. In this work we make a numerical determination of both
exponents; this is what we mean by a direct method to link correlation
length and relaxation time.

According to AG~\cite{AdamGibbs}, the barrier scales like the number
of particles involved in the rearrangement, so that 
\begin{equation}
\psi_\mathrm{AG}=d ,
\end{equation}
whereas in RFOT \cite{KTW}, the exponent $\psi$ is fixed by a
nucleation mechanism to be equal to the interfacial energy exponent,
\begin{equation}
\psi_\mathrm{RFOT} =\theta     .
\end{equation}
Notably, despite predicting quite different values of $\psi$ and
$\theta$, in both AG and RFOT relation \eqref{girobatol} reduces
to the classic Vogel-Fulcher-Tamman law of relaxation,
\begin{equation}
\tau_\mathrm{VFT}
\sim \exp \left( \frac{A}{T-T_k}\right) \ .
\label{vft}
\end{equation}

Here we find for $\psi$ and $\theta$ values different both from AG and
RFOT, but surprisingly still consistent with a plain VFT relaxation form.

\section{Numerical protocol for the study of amorphous excitations}
\label{Methods}

To estimate the two exponents $\theta$ and $\psi$ we must 
measure the energy of the amorphous excitations that 
spontaneously form and relax at equilibrium.
In particular, the exponent $\theta$ regulates the growth of the interface 
cost with 
the size of the excitation, while the exponent $\psi$ rules the 
dependence of the relaxation time on the size of cooperative regions.
Detecting amorphous excitations is not an easy task, as we lack a 
traditional order parameter (like the magnetization or the density in the
magnetic or gas-liquid first order transitions) 
able to distinguish immediately the presence of droplets of different phases.
Hence, we need a protocol to artificially build amorphous 
excitations in the system.
%%%%%%%%%%%
% add tomas
%%%%%%%%%%%
Note that the (low-temperature) excitations we are aiming to study
here are different from the \emph{dynamically} correlated regions
(dynamical heterogeneities) observed at higher temperatures, which are
not necessarily associated with an energy cost related to their size
and which may be string-shaped~\cite{schroder,glotzer}.
%%%%%%%%%%%
% end add
%%%%%%%%%%%

The core of our idea to mimic the formation of amorphous excitations is the 
following: 
we consider two independent equilibrium configurations $\alpha$ and $\beta$ 
and simply {\it exchange} all the particles contained within a sphere of 
radius $R$. In this way we directly know the size and the position of the 
excitations and the 
excess energy cost during their evolution in time.
The energy cost due to the formation of a droplet of phase $\alpha$ within a 
different phase $\beta$ (and {\it vice-versa}) can then be studied.

\begin{figure}[h]
\includegraphics[scale=0.50,angle=0]{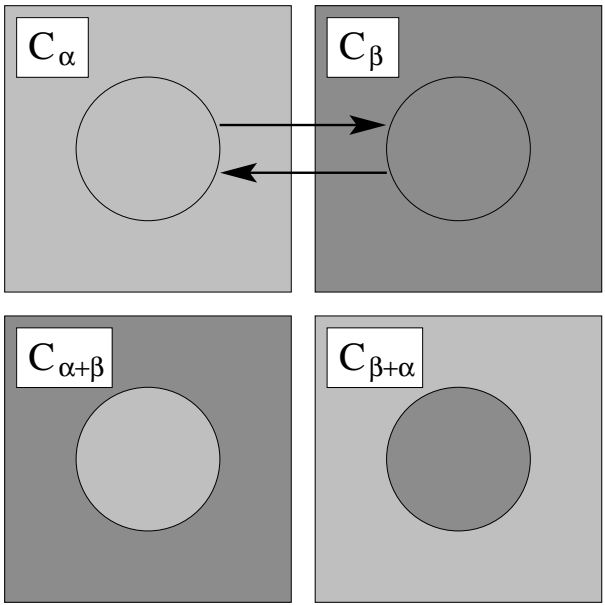}
\caption{Construction of the mixed configurations $\mathcal{C}_{\alpha+\beta}$ 
and $\mathcal{C}_{\beta+\alpha}$ using configurations $\mathcal{C}_\alpha$ and 
$\mathcal{C}_\beta$.}
\label{exchange}
\end{figure}

More in detail, this is what we do. We consider pairs of independently 
thermalized 
configurations, $\mathcal{C}_\alpha$ and $\mathcal{C}_\beta$, and from these 
we create  a mixed configuration: all particles within a sphere of fixed 
radius $R$ of $\mathcal{C}_\alpha$ are moved to a spherical 
cavity of the same shape and size
in the configuration $\mathcal{C}_\beta$; conversely, the particles within
the sphere of $\mathcal{C}_\beta$ are moved in the spherical cavity of
$\mathcal{C}_\alpha$ as it is shown in figure~\ref{exchange}. 
In this way two new configurations arise,
$\mathcal{C}_{\alpha+\beta}$ and $\mathcal{C}_{\beta+\alpha}$. 
In each initial configuration the cavity is chosen in order to preserve the 
concentration of the two species of particles in the mixed configuration 
$\mathcal{C}_{\alpha+\beta}$.
We can then study both the energetic and geometric evolution of these 
excitations at the same temperature as we used to obtain the
initial configurations. 

By using this protocol, we know exactly two very important things: 
1.\ where the excitation is;
2.\ what its geometry is, and hence what coordinate
is orthogonal to the interface between the two phases (in our case the
radial one). Such knowledge is the only reason why we are able to
compute and say something about these excitations. For spontaneous
excitations, of course, we lack both pieces of information.

This means, in particular, that we can measure the excess energy 
$\Delta E_{\alpha\beta}$ 
of the excitation, due to the interface between $\alpha$ and $\beta$,
\begin{align}
  \Delta E_{\alpha\beta} &= E_{\alpha\beta} - E^{\text{int}}_\alpha -
   E^{\text{ext}}_\beta, \nonumber \\ 
   E^{\substack{ \text{ext} \\ \text{int}}} &= \sum_{i,j: |\mathbf{r}_i| \gtrless
    R}  V_{ij}(\mathbf{r}_i-\mathbf{r}_j),
\label{DET}
\end{align}
%%%%%%%%%%%%%%
% change tomas
%%%%%%%%%%%%%%
where $V_{ij}(\mathbf{r}_i-\mathbf{r}_j)$ is the pair potential,
$E_{\alpha\beta}$ is the energy of the mixed configuration (originally
$\alpha$ inside, $\beta$ outside) and $E^{\text{int}}_\alpha$
($E^{\text{ext}}_\beta$) is the energy of the particles inside
(outside) the sphere in configuration $\alpha$ ($\beta$).
%%%%%%%%%%%%%%
% end change
%%%%%%%%%%%%%%
As we shall see, all the relevant information about the exponents $\psi$
and $\theta$ comes from the excess energy, complemented by a study of the 
geometric properties of the excitation's surface.

% Hamiltoniana

The system we consider is a binary mixture of soft particles, a fragile
glass-former~\cite{sferesoft1,SSParisi}.
In this model system, the particles are of unit mass and they 
belong to one of the two species $\gamma=1,2$, present in equal amount and
interacting via a potential: 
\beq
\mathcal{V} = \sum_{i<j}^{N}V_{ij}(|{\bf r}_{i}-{\bf r }_{j}|) =
\sum_{i<j}^{N} \biggl[ \frac{\sigma_{\gamma(i)}+\sigma_{\gamma(j)}}{|{\bf
      r}_{i}-{\bf r }_{j}|}\biggr]^{12}. 
\eeq 
The radii $\sigma_{\gamma}$ are fixed by $\sigma_{2}/\sigma_{1}= 1.2$ and
setting the effective diameter to unity, that is $(2\sigma_{1} )^{3} +
2(\sigma_{1} + \sigma_{2} )^{3} + (2 \sigma_{2})^{3} = 4 l_{0}^{3}$ , where
$l_{0}$ is the unit of length. 
The density 
is $\rho = N/V $ in units of $l_{0}$ , and we set Boltzmann's
constant $k_{B} = 1$.  A long-range cut-off at $r_{c}=\sqrt 3$ is imposed.
The thermodynamic quantities of this system depend only on 
$\Gamma=\rho/T^{1/4}$, with $T$ the temperature of the system 
\cite{book:hansen76}. 
Here $\rho=1\,l_0^{-3}$ thus the thermodynamic parameter $\Gamma$ is 
$\Gamma=1/T^{1/4}$.

The presence of two kinds of particles in the system 
%with ratio of the radii  equal to $1.2$ 
strongly inhibits the crystallization and allows the observation of
the deeply supercooled phase. Moreover an efficient Monte Carlo (MC) 
algorithm (swap
MC~\cite{GrigeraParisi}) is able to thermalize this system also at
temperature below the Mode Coupling temperature
($\Gamma_c=1.45$)~\cite{sferesoft2}.

%TEMPERATURE E NUMERO DI CAMPIONI sia per T=0 sia per T finito.
In this work we study thermalized configurations of a system with
$N=16384$ particles confined in a periodic box. The temperatures
considered, corresponding to $\Gamma=1.49$, $1.47$, $1.44$, $1.42$,
and $1.35$, are two below and three above the Mode Coupling
temperature $T_c$: $T\simeq 0.89T_c$, $0.95T_c$, $1.03T_c$, $1.09T_c$,
$1.33T_c$.
%0.897T_c, 0.947T_c, 1.028T_c, 1.087T_c, 1.331T_c
For each temperature we created mixed configurations as explained
above from a collection of about $10$ independent equilibrium
configurations, using spheres of sizes between $3$ and $8$ (in units
%%%%%%%%%%%%%%
% change tomas
%%%%%%%%%%%%%%
of $l_0$). The equilibrium configurations were produced using the swap
MC algorithm, but the relaxation of the mixed configurations was
followed using standard Metropolis MC. The upper limit of the size of
the considered excitations 
%%%%%%%%%%%%
% end change
%%%%%%%%%%%%
is due to the emergence of boundary condition effects for larger
droplets in the already numerically challenging system with $N=16384$
particles.

\section{Determination of $\psi$ and $\theta$}\label{psi}

%%%%%%%%%%%%%%
% change tomas
%%%%%%%%%%%%%%
As discussed in the introduction, the relaxation in deeply supercooled
liquids proceeds through the activated rearrangement of clusters of
correlated particles, and to these rearrangments a barrier is
associated, assumed to scale with a power of their size.  As a result,
the relaxation time is exponential in a power of the size of these
regions, eq.~\eqref{relaxation}.
%%%%%%%%%%%%%%
% change tomas
%%%%%%%%%%%%%%
In this framework we expect that the timescales involved in the
formation and in the relaxation of the cooperative regions are in fact
the same: each excitation relaxes through the cooperative
rearrangement of new excitations. This means that we can follow the
process of relaxation of an artificially produced excitation, rather
than detect the spontaneous formation of an excitation.

%%%%%%%%%%%%%%
% change tomas
%%%%%%%%%%%%%%
\begin{figure}[h]
\includegraphics[width=0.35\textwidth,angle=-90]{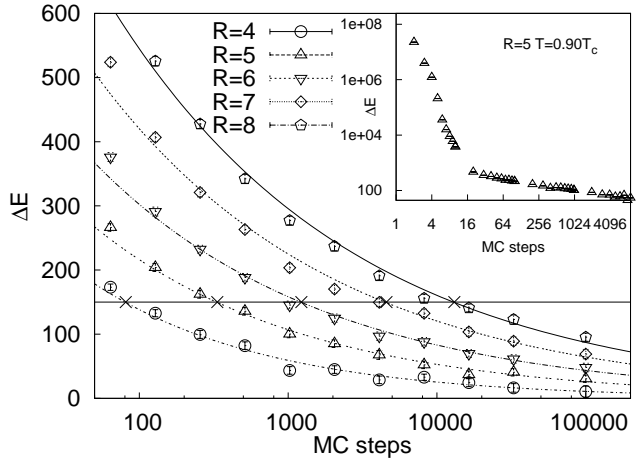}
\caption{$\langle \Delta E(R,t) \rangle$ vs.\ $t$ for $R$ ranging from
$4$ to $8$, left to right. Points are numerical data, curves are power law
$\langle\Delta E(t,R)\rangle \sim t^{-\gamma(R)} $ fits.
The fits are used to give an estimate of $\tau(R)$, represented by crosses 
on the threshold line in figure. 
The values of $\gamma$ are within the range
$0.2<\gamma(R)<0.4$. Inset: Full $\Delta E$ vs.\ $t$ time series for
$R=5$, showing the kink that separates the fast decay and the slow
relaxation regime.}
\label{stima_tau}
\end{figure}
%%%%%%%%%%%%%%
% end change
%%%%%%%%%%%%%%

%%%%%%%%%%%%%%
% change tomas
%%%%%%%%%%%%%%
\begin{figure}[h]
\includegraphics[width=0.35\textwidth,angle=-90]{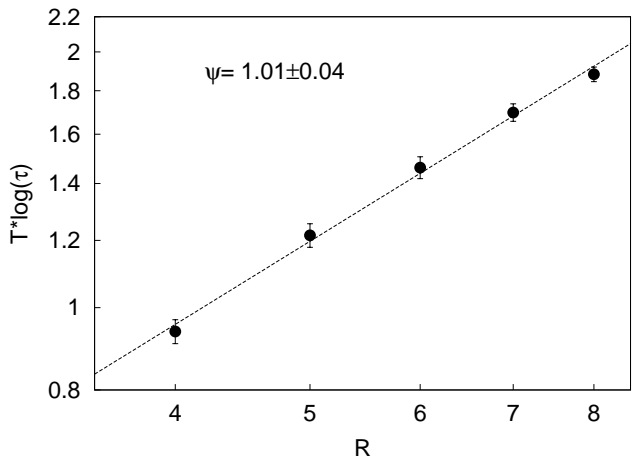}
\caption{$T \log(\tau)$ vs.\ $R$ for $T=0.89T_c$; errors are evaluated
using the
bootstrap method. The linear behaviour on log-log plot 
allows for a scaling ansatz $\tau(R)\sim \exp(R/T)$.}
\label{stima_psi}
%%%%%%%%%%%%%%
% end change
%%%%%%%%%%%%%%
\end{figure}

%%%%%%%%%%%%%%
% change tomas
%%%%%%%%%%%%%%
In practice, we study how the excess energy $\Delta E(t)$ of the
excitation relaxes with time.  The artificial construction of the
excitations has a very large effect on the excess energy only for the
first few MC steps (see Figure~\ref{stima_tau}, inset), so we
disregard the data before the kink in the $\Delta E(t)$ curve.  Figure
~\ref{stima_tau} (obtained at $T=0.89T_c$) clearly shows that the rate
of relaxation of the excitations changes with their size $R$: bigger
spheres relax over larger time scales.  Simply by fixing a threshold
value for $\Delta E(t)$, and measuring the time needed to drop below
this value, we obtain an estimate of the time $\tau$ needed to relax
an excitation of size $R$. To actually find the time at which the
threshold is reached, we need a way to interpolate between the data
points at long times. This we do by fitting a power law to the data. We 
consider the range $t>100\,$MC steps. We perform the same procedure at 
different sizes $R$ of the sphere thus obtaining a function $\tau(R)$. We 
have checked that the exponent in $\tau(R)$ curve below is insensitive to
changes in the threshold from $\Delta E_{th}=100$ to $\Delta E_{th}=175$.
%%%%%%%%%%%%
% end change
%%%%%%%%%%%%

According to~\eqref{relaxation}, $T \log(\tau)$ has to scale as
$R^\psi$. In the log-log plot in figure~\ref{stima_psi} we report $T
\log(\tau)$ {\sl vs.\/} $R$ for our lowest temperature.  The data lie with good approximation on a
straight line, thus confirming that the process of relaxation of the excitations
indeed follows the Arrhenius law. A fit of the exponent gives $\psi=1.01\pm0.04$,
and hence we conclude,
\beq
\psi= 1   .
\eeq

When a region rearranges, it is natural to expect that some excess
energy is stored at the interface between the new configuration and
the old one. The exponent $\theta$ regulates how this interfacial
energy scales with the size $R$ of the excitation,
\begin{equation}
\Delta E = Y R^\theta,
\label{nucara}
\end{equation}
where the quantity $Y$ is the (generalized) surface tension.  As
we have seen in section~\ref{intro}, the exponent $\theta$ is
crucial in order to discover the temperature dependence of the
%%%%%%%%%%%%%%%
% changes tomas
%%%%%%%%%%%%%%%
correlation length. Note however that $\theta$ is an asymptotic
exponent, and that subleading corrections to eq.~(\ref{nucara}) are in
general important for small sizes (see eq.~(\ref{ET0rough})
below). These corrections are expected from curvature (as in
liquid-liquid interfaces~\cite{review:navascues79}) or disorder
effects (as in the random field Ising model
\cite{random-field:imry75}, or the random bond Ising
model~\cite{DP}).

\begin{figure}[h]
\includegraphics[width=0.35\textwidth,angle=-90]{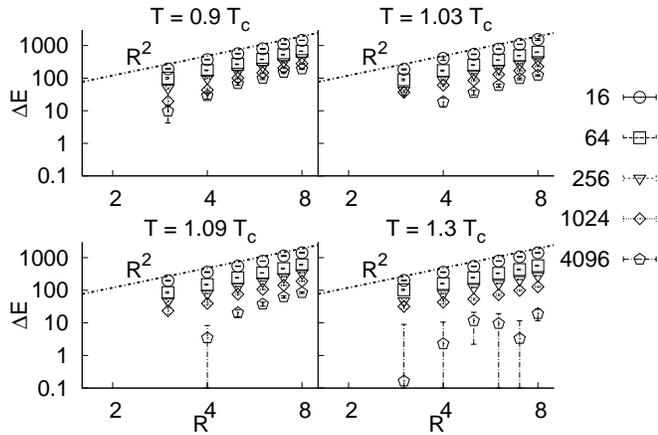}
\caption{Time dependence of the excess energy $\Delta E(t)$ vs.\ $R$ 
at four different temperatures. From left to right and top to bottom:
$T = 0.89 T_{c},1.03 T_{c},1.09 T_{c}, 1.33 T_{c}$. Time is measured 
in MonteCarlo steps.}
\label{DEtTfig}
\end{figure}

Figure~\ref{DEtTfig} shows how the excess energy $\Delta E$ defined
in~\eqref{DET} scales with $R$ for different temperatures and
times. As we have seen, the excitations decay with time, and therefore
for long times the dependence of $\Delta E$ on $R$ becomes rather
hazy. However, for intermediate times and for all analyzed
temperatures it is clear that a power law $R^2$ seems to reproduce the
asymptotic behavior rather satisfyingly. Attempting to fit the data at
long times with a single power leads to $\theta>2$, but $\theta$
cannot grow with time. Indeed, the data at very short times follow an
$R^2$ law for all sizes (as it should by construction since the
potential energy is short ranged). An exponent that grows with time is
unacceptable, because it would imply that for sufficiently large $R$,
$\Delta E$ increases with time rather than decreasing.  From this
analysis we therefore conclude
%%%%%%%%%%%%%%%
% end changes
%%%%%%%%%%%%%%%
\begin{equation}
\theta = 2  .
\end{equation}
The same value of $\theta$ was found in~\cite{articolobreve} using inherent 
structures. However, we remark that the present data are
obtained at \emph{finite} temperature, using equilibrium configurations
rather than potential energy minima. Hence, the value $\theta=2$
seems to be quite robust. Moreover, the data also show that a nonzero
surface tension $Y$ survives for quite some time after the formation
of the droplet, especially at the lowest temperatures.

\section{Consistency with the roughening exponent} \label{Measure}

Here we study the geometrical properties of the excitation
interfaces, and find further support for $\theta=2$.

In disordered systems interfaces are typically rough. The roughening
of interfaces has been investigated at length since the directed
polymer (DP) problem~\cite{DP} and the Random Field or Random Bond
Ising Model (RBIM) studies~\cite{rough,RBRF}. The signature of
roughening is the fact that the interface thickness $w$ grows with the linear size of the interface itself.
In our case it corresponds to the linear size $R$ of the excitations:
\begin{equation}
w \sim R^\gamma ,
\label{wvsL}
\end{equation}
where $\gamma$ is the so-called roughening exponent. We want to 
measure the value of $\gamma$ for the interfaces of the amorphous excitations, 
and to link it to the exponent $\theta$. However, in order to study the 
roughening properties of the excitations we must use inherent structures 
(ISs), namely minima of the potential 
energy, rather than thermal configurations as we have done up to now.
The reason is that the roughening mechanism is ruled
%%%%%%%%%%%%%%
% change tomas
%%%%%%%%%%%%%%
by a zero temperature fixed point, so that working at nonzero $T$
would needlessly introduce the complication of treating thermal
fluctuations. 
%%%%%%%%%%%%%%
% end change
%%%%%%%%%%%%%%
For the same reason, we focus on ISs obtained by equilibrium 
configurations at the lowest available temperature, $T=0.89T_c$.

The procedure to create the excitations with the ISs 
is very similar to the one described in section~\ref{Methods}.
We switch two spheres within two IS configurations 
$\mathcal{C}_{\alpha}^{IS}$  and $\mathcal{C}_{\beta}^{IS}$,
to produce the mixed configuration 
$\mathcal{C}_{\alpha+\beta}$. But such configuration is of course not
an IS itself, so we must find the new minimum 
of the potential energy, $\mathcal{C}_{\alpha\beta}^{IS}$. We do this by first  
performing $100$ Monte Carlo 
steps  at $T=0$\footnote{This first part is required because 
a single pair of particles very close together produces a huge
gradient that tends to destabilize the minimizer.} 
followed by an optimized quasi-Newton algorithm [limited memory Broyden-Fletcher-Goldfarb-Shanno (L-BFGS)~\cite{Nocedal}].
We used a collection of $16$ ISs. For each pair of configurations we used 
spheres with radii between $1.5$ and $8.5$ in units of $l_0$.

\begin{figure}[h]
\includegraphics[width=0.35\textwidth,angle=-90]{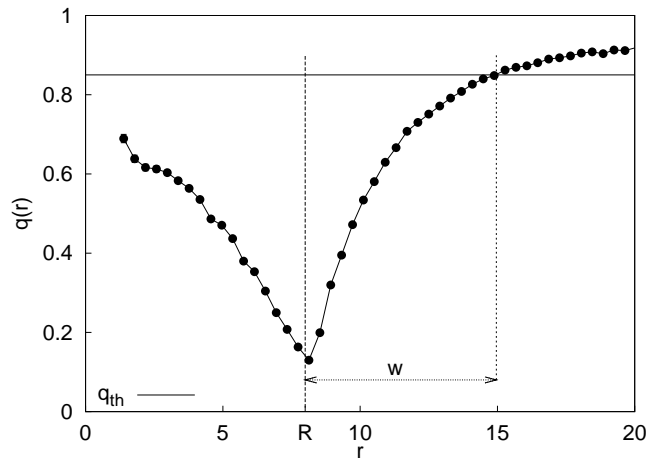}
\caption{Radial local overlap of a sphere of radius $R=8$.
For the estimate of $w$ we look only at the outer part of the interface. 
This side is not affected by finite size effects in small spheres.}
\label{overlap_locali}
\end{figure}

\begin{figure}[h]
\includegraphics[width=0.35\textwidth,angle=-90]{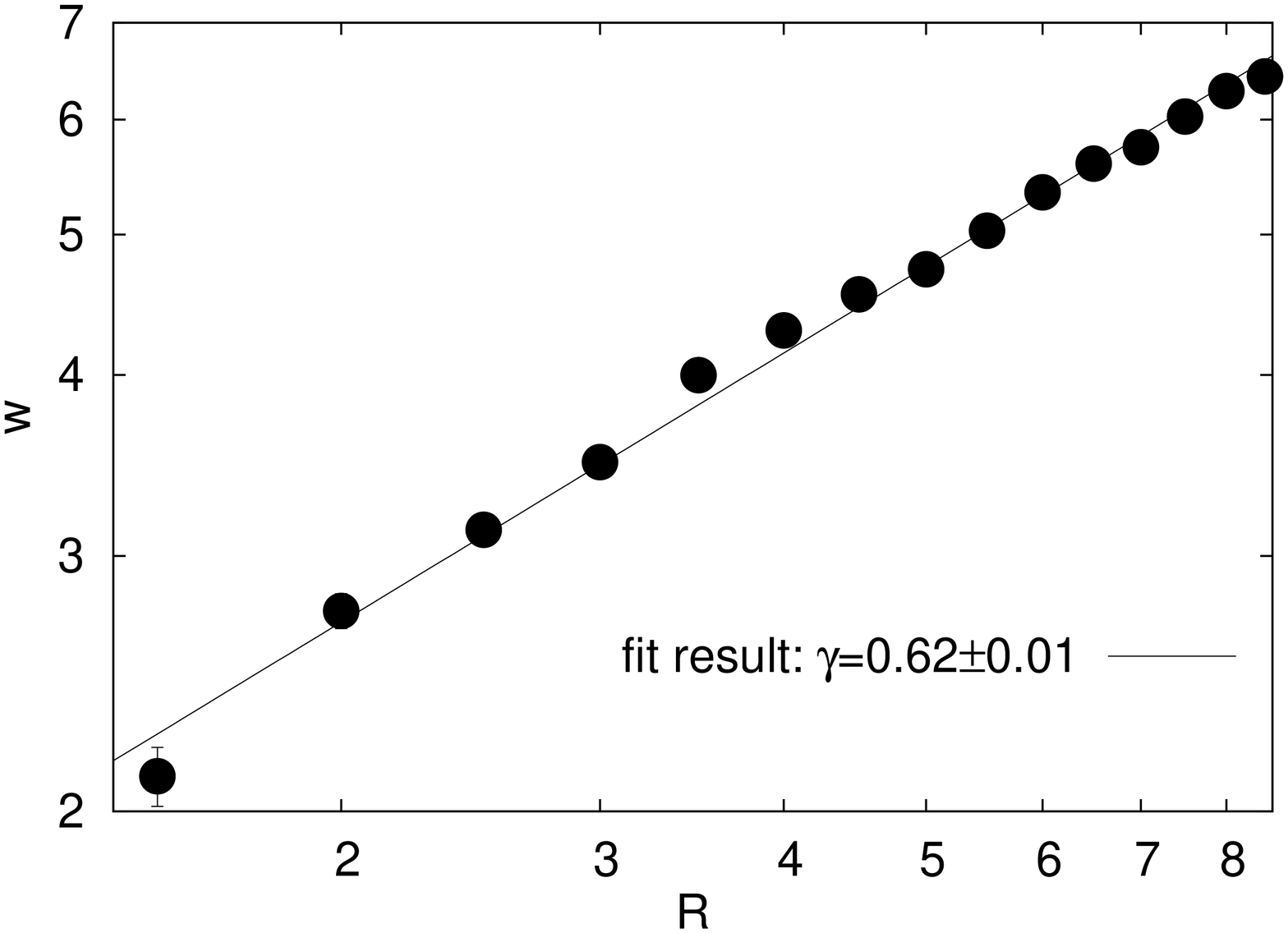}
\caption{Scaling of the interface thickness $w$ with the size $R$ of 
the sphere.}
\label{w_R}
\end{figure}

\begin{figure}[h]
\includegraphics[width=0.5\textwidth,angle=0]{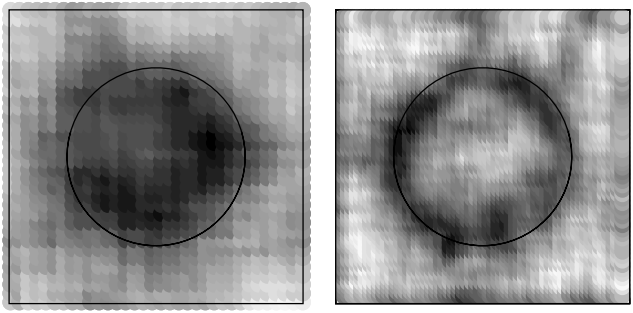}
\caption{Local overlap $q_{\alpha\beta}(r)$ between
$\mathcal{C}_{\alpha\beta}^{IS}$ and $\mathcal{C}_{\alpha+\beta}$
for configurations with interface placed at $ R = 4$ (left) and
$R=8$ (right). Plots are rescaled so that both spheres appear as
having the same size. Low overlap (dark grey) at interfaces indicate
major rearrangements of particles, while far from interfaces the
overlap with initial configurations is high (light grey). Clearly
the larger sphere (right) has a relatively thinner, or smoother,
interface.}
\label{overlap}
\end{figure}

In order to define the thickness $w$ of the excitation we can look at 
the overlap $q(r)$ ($r$ is the distance from the origin)
between the just-switched configuration $\mathcal{C}_{\alpha+\beta}$
and its relative minimum $\mathcal{C}_{\alpha\beta}^{IS}$ 
(for a definition of the local overlap see 
appendix~\ref{overlapapp}). The overlap is small close to the 
interface, where particles have moved most, while it gets close to one
away from it. Hence, 
we can define $w$ as the thickness of the region for which the local 
radial overlap of the excitation is smaller than an arbitrary threshold value 
$q_{th}$.
In figure~\ref{overlap_locali}, the local overlap is plotted as a function of
the distance $r$ from the center of the sphere.
Since in very small spheres the overlap has no room to reach high enough 
values in the inside, we actually define $w$ as sketched in 
figure~\ref{overlap_locali}.
In figure~\ref{w_R}, we report  $w$ as a function of the radius $R$ 
of the excitation
in a log-log plot. We conclude that the excitations' interfaces roughen 
according to relation
\eqref{wvsL}, with
\begin{equation}
\gamma=0.62\pm0.01.
\label{blasco}
\end{equation}
A roughening exponent smaller than one implies that the ratio $w/R$
between width and linear scale of the surface decreases for larger
spheres.  Thus large excitations have \emph{relatively} thin
interfaces.  This is clearly shown in figure~\ref{overlap}, where two
excitations with different radius ($R=4$ vs. $R=8$) are compared in a
coordinate system where all lengths are rescaled by $R$, so that both
rescaled spheres have virtual radius unity. The rescaling emphasizes
the thick interface of the smaller excitation compared to the sharper
interface of the larger excitation.

To understand why roughening occurs we can think about the
Random Bond Ising Model (RBIM).
This is an Ising spin model where the nearest neighbours bonds are random, 
albeit typically positive. 
In such a system the position of a domain wall strongly depends on the disorder,
since the weak bonds are more likely to be broken.
On one hand, a smooth domain wall is preferable, as it would break the 
smallest number of bonds. 
On the other hand, some suitable deviation from smoothness could induce the 
breaking of weaker bonds and hence a lower energy cost.
Hence, a rough interface is the result of a complicated optimization problem: 
the cost of a large number of broken bonds is balanced by the gain due to the 
presence of very weak bonds among them. As a result, in a disordered system a 
rough interface can be 
energetically favoured with respect to a smooth interface.

The interesting point is that in the context of elastic manifolds in
random media~\cite{SAD,KSA}, a precise relation exists between the
energy gain we just mentioned and the roughening exponent $\gamma$,
and in such relation the exponent $\theta$ comes into play.  The
energy gain by roughening appears as a negative correction to the
ground state energy of the manifold,
\begin{equation}
E_s=Y R^\theta-B R^{2\gamma},   \qquad d=3  .
\label{ET0rough}
\end{equation}
Our hypothesis is that the interface energy $\Delta E_{\alpha\beta}$
of the amorphous excitations can be described as in \eqref{ET0rough}
by a leading term (due to a generalized surface tension) plus a
sub-leading correction due to roughening. In random manifolds
$\theta=d-1$, whereas we do not know $\theta$. We thus try to find
$\theta$ by taking the value of $\gamma$ found in~\eqref{blasco} and
using eq.~\eqref{ET0rough} to fit $\Delta E_{\alpha\beta}$ with
$\theta$ as a fitting parameter.

\begin{figure}[h]
\includegraphics[width=0.35\textwidth,angle=-90]{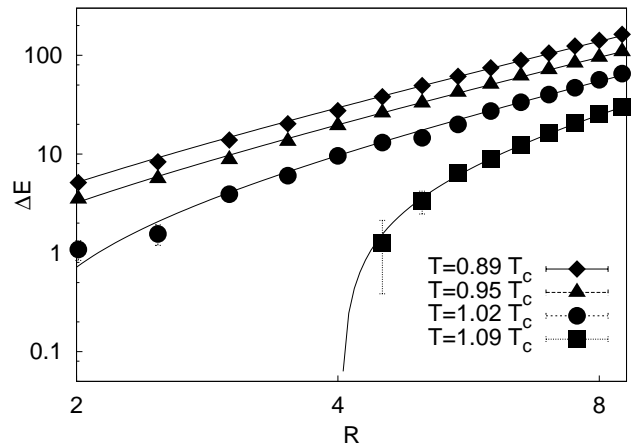}
\caption{$\langle\Delta E^{IS}_{\alpha\beta}(R)\rangle$ behaviour with $R$ at 
different temperatures $T$ ranging from $0.89 T_c$ to $1.33 T_c$. 
The log-log plot enhances the power law trend at large $R$. Lines are
fits to eq.~(\ref{ET0rough}) with $\theta$ and $\gamma$ fixed (see text).}
\label{de_R_inset}
\end{figure}

The interface energy $\Delta E_{\alpha\beta}$ is calculated as in
eqs.~(\ref{DET}), except that $E_{\alpha\beta}$ is the energy of
$\mathcal{C}_{\alpha\beta}^{IS}$.  In figure~\ref{de_R_inset} we plot
$\Delta E_{\alpha\beta}$ vs.\ $R$ at different temperatures.  A
deviation from a simple power law at small sizes is evident.  A fit of
$\Delta E_{\alpha\beta}$ at the lowest temperature $T=0.89T_c$ with
the functional form~\eqref{ET0rough} (not shown) and $\gamma=0.62$
gives \beq \theta=2.08\pm0.04 .  \eeq
%%%%%%%%%%%%%%
% change tomas
%%%%%%%%%%%%%%
Reversing the procedure (i.e.\ fixing $\theta=2$ and fitting $\gamma$
in using~\eqref{ET0rough}) gives $\gamma=0.75$.  Finally, fixing both
$\theta=2$ and $\gamma=0.62$ and $\theta=2$, still gives an excellent
fit of $\Delta E_{\alpha\beta}$ vs.\ $R$ (figure~\ref{de_R_inset}).
Since we know that $\theta$ cannot exceed 2, this consistency check
confirms the value of $\theta$ found before.
%%%%%%%%%%%%%%
% change tomas
%%%%%%%%%%%%%%

\section{Discussion}

Our result $\theta=2$ is somewhat sensible and not particularly
exciting: it is basically telling us that disorder in a supercooled
liquid is not strong enough to change in any exotic way the leading
term of the surface energy cost: surfaces remain surfaces, albeit a
bit rough. $\theta$ is not actually a part of AG theory, so it is more
proper to compare with RFOT, where $\theta$ was originally introduced.
It must be said that the value $\theta_\mathrm{RFOT}=d/2$ derived in
\cite{KTW} using RG arguments always had its greatest appeal in the
fact that, together with $\psi_\mathrm{RFOT}=\theta_\mathrm{RFOT}$, it
gave back the VFT equation~\eqref{vft}. In fact, the arguments used in
\cite{KTW} to fix $\theta$ do not belong to RFOT itself, and other
values are in principle compatible with the conceptual structure of
RFOT. On the contrary, the value found for $\psi$ within RFOT has to
do with a crucial aspect of the theory.  Hence, it seems that the real
interesting comparison is about the exponent $\psi$, after all.

The value $\psi=1$ we find implies that the barrier for the
rearrangement of a correlated region scales linearly with 
its size,
\begin{equation}
\Delta \sim \xi   .
\label{utauta}
\end{equation}
The proposal of AG, $\Delta\sim \xi^d$, has always seemed a bit
exaggerated, since one can imagine several ways for the system to pay
\emph{less} than the entire volume to rearrange a region.
 One of this
ways, and not the most pedestrian, is suggested by the fact that the
greatest part of the energy necessary to create the excitation is
stored in the interface. Hence, one may expect that the barrier scales
with the same exponent as the surface energy cost, i.e.\
$\theta$. This is, basically, the idea of RFOT. We note that such idea
is deeply rooted in the theory of nucleation, which was a indeed a
source of inspiration for the original formulation of RFOT.
%%%%%%%%%%%
% add tomas
%%%%%%%%%%%
A proportional relationship between inverse diffusion
constant and the exponential of the number of particles belonging to a
dynamically correlated cluster has been reported for a model of
water~\cite{giovambattista}, but this is not necessarily in
contradiction with the above. The dynamical clusters and the
excitations of RFOT (or AG's cooperatively rearranging regions) are
probably different entities, as mentioned above. Furthermore, the
simulations of ref.~\onlinecite{giovambattista} correspond to a
different temperature regime (above the mode-coupling temperature) and
the objects found in that work are rather noncompact, with different
volume/surface ratio from the excitations we are studying.
%%%%%%%%%%%
% end add
%%%%%%%%%%%

According to RFOT there are two competing forces: the free-energy cost
to create an excitation, scaling as
$Y R^\theta$, and the configurational entropy gain due to the change
of state of the rearranging region, scaling as
$TS_c R^d$. In these two expressions $Y$ is the surface tension and
$S_c$ is the configurational entropy. Hence,
the total free energy for the formation of the excitation is, according to RFOT,
\begin{equation}
\Delta F (R)= Y R^\theta - TS_c R^d     .
\label{nuoro}
\end{equation}
At this point, RFOT, in perfect analogy with
nucleation theory, proceeds by finding the {\it maximum} of such
non-monotonous function (recall that $\theta < d$). This maximum
provides two
essential pieces of information: First, the position of the maximum,
$R=\xi$, gives the critical size of the rearranging region, i.e.\ the
mosaic correlation length (cf.\ eq.~\ref{eq:1}),
\begin{equation}
\xi =\left(\frac{Y}{TS_c}\right)^\frac{1}{d-\theta}      .
\label{marro}
\end{equation}
Second, and most important for us now, the height of the maximum,
$\Delta F(R=\xi)$, gives the size of the free-energy
barrier to be crossed to rearrange the region,
\beq
\Delta \equiv \Delta F(R=\xi) \sim \xi^\theta     .
\eeq
This fixes $\psi_\mathrm{RFOT}=\theta_\mathrm{RFOT}$, and it
coincides with the intuitive notion that the barrier
should scale the same as the interface cost.

This last result, however, is at variance with what we find here,
relation \eqref{utauta}: in fact, whatever one thinks about our
numerical result for $\theta$, a value of $\theta$ as small as $1$
seems rather unlikely. In any case, it is important to emphasize that
$\psi=\theta$ is a consequence of the \emph{maximization} of
eq.~\eqref{nuoro}, which in turn follows from the nucleation
paradigm. It has been noted, however, that nucleation is perhaps not a
fully correct paradigm to describe the formation of amorphous
excitations within a deeply supercooled liquid \cite{BB}.  The essence
of RFOT, namely the competition between a surface energetic term and a
bulk entropic term, retains its deepest value even if we do not cast
it within the strict boundaries of nucleation theory. The value of the
correlation length may come from the point where the two contributions
balance, rather than from the maximum of \eqref{nuoro}, and (for
obvious dimensional reasons) one gets the same expression
\eqref{marro} for $\xi$ (up to an irrelevant constant), while $\psi$
remains undetermined. These points are discussed in depth in~\cite{BB}. 
Here we simply note that our present results are quite
compatible with RFOT in the form it has been recast in~\cite{BB},
without reference to a nucleation mechanism.

%%%%%%%%%%%%
% add andrea
%%%%%%%%%%%%
Regarding the comparison between $\theta$ and $\psi$ there is a final
point we have to discuss. The reader familiar with spin-glass physics
will probably remember the Fisher-Huse (FH) inequality, \cite{fh},
\beq
\psi \geq \theta  ,
\label{fh}
\eeq
which is plainly violated by the values we find here. The physical
motivation of \eqref{fh} is basically the following: if we represent
the excitation as an asymmetric one dimensional double well, where the
abscissa is the order parameter and the ordinate is the energy of the
excitation, the height of the barrier (which scales as $\xi^\psi$) is
always larger than (or equal to) the height of the secondary minimum
(which scales as $\xi^\theta$), and hence $\psi\geq\theta$.

How comes, then, that we find $\psi < \theta$? The FH argument was
formulated in the context of the droplet picture for spin-glasses,
where there are only {\it two} possible ground states.  In supercooled
liquids, in contrast, a rearranging region can choose among an {\it
  exponentially large} number of target configurations.  Under this
conditions, even though the FH bound still applies to the {\it energy}
barrier, there may be a nontrivial entropic contribution that
decreases the \emph{free energy} barrier to rearrangement. We
determine $\psi$ by measuring a time, and hence a free-energy barrier,
not an energy barrier, so that the FH constraint does not necessarily
hold in the case of supercooled liquids.  This entropic effect is
absent in the original FH argument due to the lack of exponential
degeneracy of the target configurations (in fact, one would expect
relation \eqref{fh} to hold even in the mean-field picture of
spin-glasses, where the number of ground states is large but the
configurational entropy is still zero). How the FH argument should be
modified in the presence of such large entropic contribution is
however not clear at this point.
%%%%%%%%%%%%
% end add
%%%%%%%%%%%%

\section{Conclusions}

We have determined numerically (in $d=3$) the exponents linking the
size of rearranging regions to barrier height ($\psi$) and to surface
energy cost ($\theta$). This is to our knowledge the first direct
measure of these exponents. We find
\begin{equation}
\psi = 1, \qquad \theta=2.
\end{equation}
Both values are in disagreement with those assumed by the AG and RFOT
schemes. However, if we stick to eq.~\eqref{girobatol}, we still
obtain for the relaxation time the VFT relation~(\ref{vft}).  It seems
that, albeit changing all cards on the table, we managed to get back
the most used fitting relation in the physics of glass-forming
systems. We stress that there is no particular reason to stick to
such VFT In fact, as it has been remarked many times before, a
generalised VFT with an extra fitting exponent $\nu$, such that $\log
\tau \sim (T-T_k)^{-\nu}$, would do an even better job in fitting the
data. Yet, to get back VFT as the product of the independent numerical
determination of two rather different exponents, remains a rewarding
result to some extent.

Finally, let us remark that the disagreement between our exponents and
those proposed in the original RFOT do not imply as harsh a blow to
RFOT as it might seem at first. RFOT can be cast in a form~\cite{BB}
that retains its most essential aspect, namely the competition between
a surface energy cost and a bulk energy gain, without using nucleation
theory. If one does this, the exponents $\psi$ and $\theta$ remain
unrelated and compatible with our findings.  Within this context, our
result $\psi=1$ seems to be an indication that RFOT is a better theory
if one does not push the nucleation analogy too hard.

\section{Acknowledgements}

The authors thank G. Biroli, S. Franz, and I. Giardina
for interesting discussions, 
and ETC* and CINECA for computer time. 
The work of TSG was supported in part by
grants from ANPCyT, CONICET, and UNLP (Argentina).

\begin{appendix}

\section{Definition of the local overlap}
\label{overlapapp}
A suitable definition of the
overlap, for the off-lattice system considered, is given using a method
similar to that used in \cite{PRL}:
we divide the system in $64^3$ small cubic boxes with side
$L/64$ and, having two configurations $\sigma$ and $\tau$,
we compute the quantity
\beq
q_{\sigma\tau}(x,y,z)=n_\sigma(x,y,z) n_\tau(x,y,z)\ ,
\label{qbox}
\eeq
where $n_\sigma(x,y,z)$ is $1$ when the box with center at coordinates $x,y,z$
contains at least one particle and it is $0$ when the same box is empty.
To each box we assign the weight
\beq
w_{\sigma\tau}(x,y,z)=\frac{n_\sigma(x,y,z) + n_\tau(x,y,z)}{2}\ .
\label{wbox}
\eeq
The global overlap $q_{\sigma\tau}$
between these two configurations $\sigma$ and $\tau$ is given by
%using the sum $\sum_{}$ over all the boxes of the system:                                               
\beq
q_{\sigma\tau}=\frac{\sum_{i} q_{\sigma\tau}(x_i,y_i,z_i)
w_{\sigma\tau}(x_i,y_i,z_i)}{\sum_{i} w_{\sigma\tau}(x_i,y_i,z_i) }\ .
\label{qTOT}
\eeq
where $i$ runs over all boxes in the system.

We are interested in the local value of the overlap $q_{\alpha\beta}(r)$
at distance $r$ from the centre of the sphere.
The definition of the local overlap $q_{\alpha\beta}(r)$ for a spherical corona 
between $r-dr$ and $r+dr$ is given by considering in \eqref{qTOT}
only the sum of boxes belonging to this region.

\end{appendix}

%\listoffigures

%\newpage

%\thispagestyle{plain}

\end{document}